\documentclass[twocolumn,showpacs,preprintnumbers,prl]{revtex4}
\usepackage{physics}
\usepackage{graphicx}
\usepackage{dcolumn}
\usepackage{bm}
\usepackage{color}

\begin{document}

\title{Manipulation of Multipartite Entanglement in an Array of Quantum Dots}

\author{Yen-Ju Chen}
\email{yjchen@m105.nthu.edu.tw}
\author{Chih-Sung Chuu}
\email{cschuu@phys.nthu.edu.tw}
\affiliation{Department of Physics, National Tsing Hua University, Hsinchu 30013, Taiwan\\
Center for Quantum Technology, Hsinchu 30013, Taiwan}

\begin{abstract}
Multipartite entanglement is indispensable in the implementation of quantum technologies and the fundamental test of quantum mechanics. Here we study how the W state and W-like state may be generated in a quantum-dot array by controlling the coupling between an incident photon and the quantum dots on a waveguide. We also discuss how the coupling may be controlled to observe the sudden death of entanglement. Our work can find potential applications in quantum information processing.
\end{abstract}


\maketitle

\section{Introduction}

Quantum entanglement has been at the heart of implementing quantum technologies and understanding quantum mechanics, with examples ranging from quantum cryptography \cite{Ekert91, Bennet92} and quantum teleportation \cite{Bennett93} to Bell's theorem \cite{Bell64}. When the number of the entangled particles or systems exceeds two, the multipartite entanglement enables striking applications such as the measurement-based quantum computation \cite{Raussendorf01, Knill01}, quantum error correction \cite{Shor94}, quantum secret sharing \cite{Hillery99}, and quantum simulation \cite{Lloyd96} as well as the understanding of the transition from quantum to classical regime \cite{Leggett08}. So far, the entanglement of fourteen ions \cite{Monz11}, ten photons \cite{Wang16}, and ten superconducting qubits \cite{Song17} have been demonstrated experimentally. The multipartite entanglement of quantum dots (QDs), by comparison, is more challenging to achieve. The photon-electron entanglement has been demonstrated in a single QD \cite{Greve12, Gao12} and the entanglement of two QDs has also been reported \cite{Delteil16}.

In this paper, we explore how entanglement may be generated in a QD array by manipulating the coupling between the optically active QDs and an incident photon of controllable waveforms on a waveguide. The multipartite entanglement discussed here includes the W and W-like states \cite{Dur00}, which are known for its robustness against the qubit loss and central for the optimal universal and state-dependent quantum cloning \cite{Burb98}, respectively. In addition, we also study how the entanglement of arbitrary QD pairs in the array can be controlled to observe the sudden death of entanglement~\cite{Yu09}. To the best of our knowledge, the generation of W-like state and observation of sudden death have never been discussed in the quantum-dot system. The generation of the W states has been proposed using the spin states with bulky optics \cite{Kang15}, with two-qubit gates, Kerr nonlinearity, and photon-number-resolving detection \cite{Heo19}, with multiple quantum dots in a cavity \cite{Liu20}, or with the quantum-dot molecules \cite{Liu14}. In comparison, the scheme proposed here is based on the exciton states and is suitable for scaling up due to its simpler implementation. A possible candidate of the QDs considered in this work is the chemically synthesized CdSe QDs \cite{Michler00} coupled to a plasmonic waveguide with the nanopatterning technology \cite{Manfrinato13, Xie15, Chen18}. At low temperature, the radiative recombination (via phonon emission or absorption) from the exciton ground state $\ket{e}$ in these QDs is much slower than the direct optical recombination to the zero exciton state $\ket{g}$ (via the mixing with the lowest-energy optically active exciton state), thus resulting in a long-lived upper state $\ket{e}$ (radiative lifetime of $\sim 2 \ \mu$s at 2.3 K) and lower state $\ket{g}$ \cite{Labeau03}.
  
%the generation of collective states in an atomic ensemble with $\lambda$-type energy levels has been proposed in the DLCZ protocol \cite{Duan01}. This method has also been extended to generate collective atomic states with definite numbers of excitations deterministically or probabilistically using the long-range dissipative or coherent coupling induced via a waveguide \cite{Gonzalez15a,Paulisch17,Gonzalez13,Gonzalez14,Gonzalez15b}. In comparison, ancilla qubits and the $\lambda$-type energy level are not required in our work.

%This paper is organized as follows. In Sec.~II we first introduce our theoretical model. The generation and manipulation of the multipartite entanglement (the W and W-like states) are then presented in Sec.~III. Finally, other possibilities of manipulating the entanglement (for example, the generation and sudden death of the entanglement between arbitrary QD pair) are discussed in Sec.~IV before we conclude our work in Sec.~V.

\section{Theoretical model}

\begin{figure}[ht!]
\centering
\includegraphics[width=8.5cm]{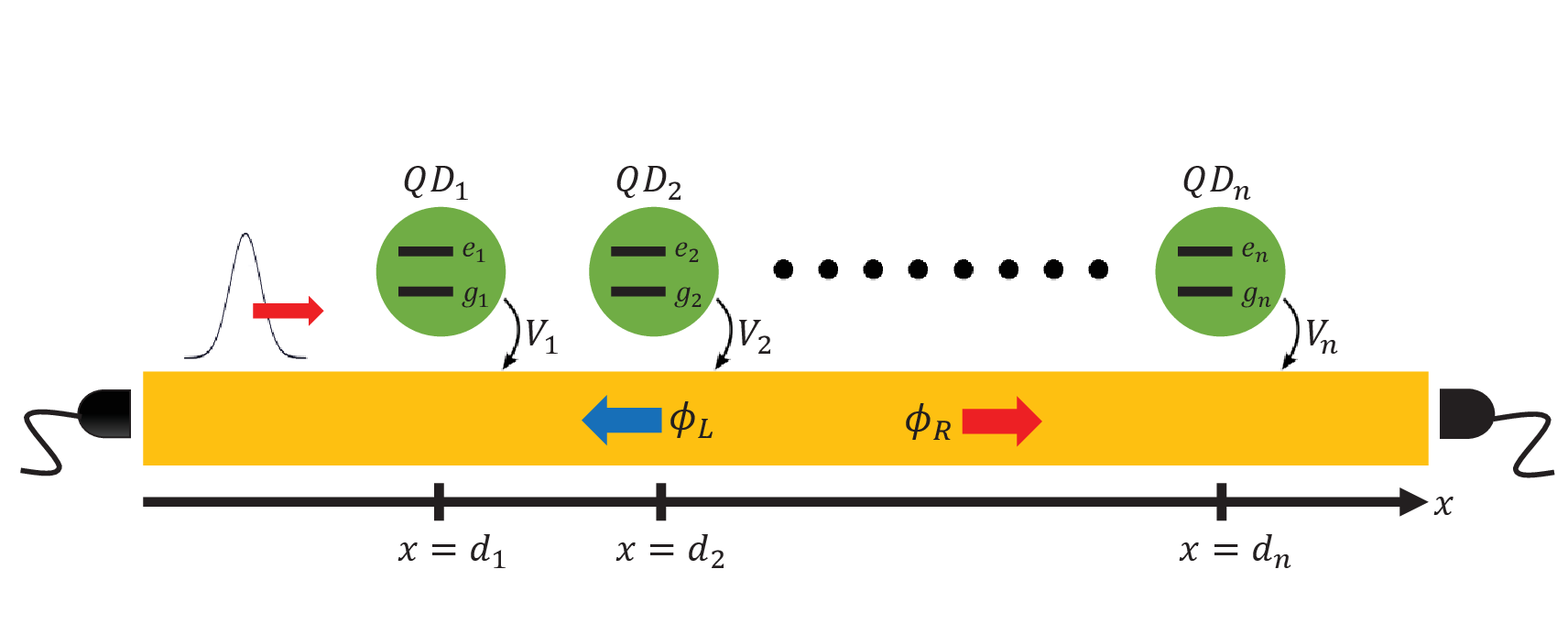}
\caption{\label{fig:1} A chain of two-level quantum dots QD$_i$ ($i=1,2,3, ...$) located at $x_i$ are coupled to a waveguide with coupling strength $V_i$ and detuning $\delta_i$. The entanglement between the QDs can be generated by a single photon incident from the left with properly tuned $V_i$ and $\delta_i$.}
\end{figure}

The proposed scheme for generating multipartite entanglement is illustrated in Fig.~\ref{fig:1}, where $N$ two-level QDs at $x=d_{1}, d_{2}$ ... $d_{N}$ are positioned nearby a plasmonic waveguide. Suppose a single photon is incident from the left of the waveguide. The real-space Hamiltonian of the coupled system, which can also be used for studying the single-photon transport~\cite{Shen05,Chang07,Shen09,Cheng16,Cao17}, is given by
\begin{equation}
\begin{aligned}
	H = & \, \hbar \int{dx \left[-iv_{g} C_{R}^{\dag}(x) \frac{\partial}{\partial x} C_{R} (x)  
													+iv_{g} C_{L}^{\dag}(x) \frac{\partial}{\partial x} C_{L} (x) \right]}\\
		 &+ \sum_{j=1}^N \hbar \int{dx \,V_{j}\delta(x-d_{j}) [ C_{R}^{\dag}(x)\sigma_{j} + C_{R}(x)\sigma_{j}^{\dag}} \\
																												&+C_{L}^{\dag}(x)\sigma_{j} + C_{L}(x)\sigma_{j}^{\dag}]
	   + \sum_{j=1}^N \hbar \left( \omega_{j}-i \frac{\Gamma_{j}}{2} \right) \sigma_{j}^{\dag} \sigma_{j},
\end{aligned}
\end{equation}
%\begin{eqnarray}
%	H &=& \hbar \int {dx \left[-iv_{g} C_{R}^{\dag}(x) \frac{\partial}{\partial x} C_{R} (x) + iv_{g} C_{L}^{\dag}(x) \frac{\partial}{\partial x} C_{L} (x) \right]} \nonumber \\
%		 &+& \sum^N_{j=1} \hbar \int{dx \,V_{j}\delta(x-d_{j}) [ C_{R}^{\dag}(x)\sigma_{j} + C_{R}(x)\sigma_{j}^{\dag} \nonumber \\
%	   &+& C_{L}^{\dag}(x)\sigma_{j} + C_{L}(x)\sigma_{j}^{\dag}]} + \sum_{j=1}^N \hbar \left( \omega_{j}-i \frac{\Gamma_{j}}{2} \right) \sigma_{j}^{\dag} \sigma_{j}
%\end{eqnarray}
where $C_{R}^{\dag} (x)$ and $C_{L}^{\dag} (x)$ are the bosonic operators creating a right- and left-moving photons at position $x$, respectively. $\sigma_{j}^{\dag}=\ket{e_{j}}\bra{g_{j}}$ and $\sigma_{j}=\ket{g_{j}}\bra{e_{j}}$ are the raising and lowering operators for the $j$-th QD, respectively. $v_{g}$ is the group velocity of the photon in the waveguide. $\omega_{j}$ is the transition frequency of the $j$-th QD. $\Gamma_{j}$ describes the energy dissipation rate of the $j$-th QD, including the free space spontaneous decay and other decay channels (for example, the Ohmic loss in a plasmonic waveguide). $V_{j}$ is the coupling strength between the $j$-th QD and the waveguide. 

The eigenstates in the single-excitation subspace with energy $E=v_{g} k$ equal to the propagating photons can then be written as \cite{Shen09}
\begin{equation}
\begin{aligned}
 \ket{E}&=\int{dx \left[ \phi_{R}(x) C_{R}^{\dag}(x) + \phi_{L}(x) C_{L}^{\dag}(x) \right]} \times\\
					&\ket{g_{1},\cdots,g_{N}}\ket{0}_{\rm wg} + \sum_{j=1}^N \xi_{j}\sigma_{j}^{\dag} \ket{g_{1},\cdots,g_{N}}\ket{0}_{\rm wg},
\end{aligned}
\end{equation}
%\begin{eqnarray}
% \ket{E}&=&\int{dx \left[ \phi_{R}(x) C_{R}^{\dag}(x) + \phi_{L}(x) C_{L}^{\dag}(x) \right] \times \nonumber \\
%				&&\ket{g_{1},\cdots,g_{N}}\ket{0}_{\rm wg}} + \sum_{j=1}^N \xi_{j}\sigma_{j} \ket{g_{1},\cdots,g_{N}}\ket{0}_{\rm wg}
%\end{eqnarray}
where $\phi_{R}(x)$ and $\phi_{L}(x)$ are the wave functions of the right- and left-moving photons in the waveguide, respectively. $\xi_{j} = \xi'_j{\rm exp} (\phi_{\xi'_j})$ is the probability amplitude, with $|\xi'_j|^2$ and $\phi_{\xi'_j}$ being the excitation probability and phase of the $j$-th QD. $\ket{g_{1},\cdots,g_{N}}\ket{0}_{\rm wg}$ denotes that all QDs are in their ground states with no photon in the waveguide.                                                                                                                                  

Now, suppose a single-photon detector is placed at each end of the waveguide. If neither detector clicks, the quantum state of the QDs is projected onto 
\begin{equation}
	\ket{\psi} = \sum_{j=1}^N \xi'_{j} {\rm exp} (i\phi_{\xi'_{j}}) \sigma_{j}^{\dag}\ket{g_{1},\cdots,g_{N}}
	\label{eq:multipartite}
\end{equation}
with a probability of $\sum_{j} |\xi'_{j}|^2$, which can be maximized by shaping the single photons \cite{Zhang12} or reducing the length of the plasmonic waveguide by interconnecting each section of the coupled QD and plasmonic waveguide by dielectric waveguide. As we will show below, this multipartite entangled state can be manipulated by controlling the coupling strength and detuning of the QDs, which may be achieved experimentally via tuning the distance between the QD and waveguide, the separation between adjacent QDs, or the strain and electric field on the QDs.

%The transport properties of single photon scattered by three QEs have been discussed in ref.\cite{Huang18}. By further controlling the azimuthal angles of each QEs appropriately, the desired transmission and reflection can be achieved, which may have potential applications such as single photon switch.

\section{Multipartite Entanglement}

For the sake of simplicity, we will consider three QDs coupled to the plasmonic waveguide. The wave functions $\phi_{R}(x)$ and $\phi_{L}(x)$ then take the forms
\begin{equation}
\begin{aligned}
  \phi_{R}(x)=&e^{ikx}[\,\theta(d_{1}-x)+a_{1}\theta(x-d_{1})\theta(d_{2}-x)  \\
	 										&+a_{2}\theta(x-d_{2})\theta(d_{3}-x)+t\theta(x-d_{3})], \\
	\phi_{L}(x)=&e^{-ikx}[\,r\theta(d_{1}-x)+b_{1}\theta(x-d_{1})\theta(d_{2}-x) \\
	 										&+b_{2}\theta(x-d_{2})\theta(d_{3}-x)].
\end{aligned}
\end{equation}
Here, $t$ and $r$ are the transmission and reflection amplitudes of the single photon, respectively. $\theta(x)$ is the Heaviside step function with $\theta(0)=1/2$. $a_i$ and $b_i$ are the probability amplitudes of a right- and left-moving photon, respectively, in between $x = d_i$ and $d_{i+1}$. Using these wave functions and the eigenvalue equation $H\ket{E}=E\ket{E}$, the following relations of the probability amplitude, coupling strength and detuning can be obtained,
\begin{equation}
\begin{aligned}
	&\left(M_{11}-\delta_{1}-i\frac{\Gamma_{1}}{2}\right)\xi_{1} 
	+ M_{21}\xi_{2}
	+ M_{31}\xi_{3}
	= -V_{1}e^{ikd_1}, \\
	& M_{21}\xi_{1}
	+ \left(M_{22}-\delta_{2}-i\frac{\Gamma_{2}}{2}\right)\xi_{2} 
	+ M_{32}\xi_{3}
	= -V_{2}e^{ikd_2}, \\
	&	M_{31}\xi_{1}
	+ M_{32}\xi_{2}
	+ \left(M_{33}-\delta_{3}-i\frac{\Gamma_{3}}{2}\right)\xi_{3} 
	= -V_{3}e^{ikd_3},
\end{aligned}
\label{eq:relations}
\end{equation}
where $M_{jl}=-i(V_{j}V_{l}/v_{g}){\rm exp}[ik(d_{j}-d_{l})]$ and $\delta_{j}=E/\hbar-\omega_{j}$ is the detuning of the single photon relative to the transition frequency of the $j$-th QD. 

%\begin{figure}[ht!]
%\centering
%\includegraphics[width=8cm]{fig2}
%\caption{\label{fig:2} (a) Excitation probabilities and (b) phases of the W state as a function of the detuning $\delta_1$. The detuning is in the unit of the decay rate into the waveguide mode $\Gamma_{\rm wg}=V^{2}/v_{g}$. In the calculation, $V_{1}=V_{2}=V_{3}$, $\delta_{1}=\delta_{2}=\delta_{3}$, $k(d_{2}-d_{1})=k(d_{3}-d_{2})=2\pi$ and $\Gamma_{1}=\Gamma_{2}=\Gamma_{3}=0$.}
%\end{figure}

To investigate the possibility of generating the W or W-like states, we first take equal coupling strength ($V_1=V_2=V_3$) and detuning ($\delta_1=\delta_2=\delta_3$), and place the QDs by a spacing of multiple photon wavelengths, i.e. $k(d_{2}-d_{1})=k(d_{3}-d_{2})=2 m \pi$, where $m$ is an integer. The dissipation is neglected at this moment. Eq.~(\ref{eq:relations}) then gives $|\xi'_1|^2=|\xi'_2|^2=|\xi'_3|^2$ and $\phi_{\xi'_1}=\phi_{\xi'_2}=\phi_{\xi'_3}$ independent of the detuning. Thus, the W state, 
\begin{equation}
  \ket{W}=\frac{1}{\sqrt{3}}\left(\ket{e_{1},g_{2},g_{3}}+\ket{g_{1},e_{2},g_{3}}+\ket{g_{1},g_{2},e_{3}}\right),
	\label{eq:W}
\end{equation}
can be generated in the QD chain up to a global phase.

\begin{figure}[ht!]
\centering
\includegraphics[width=8.5cm]{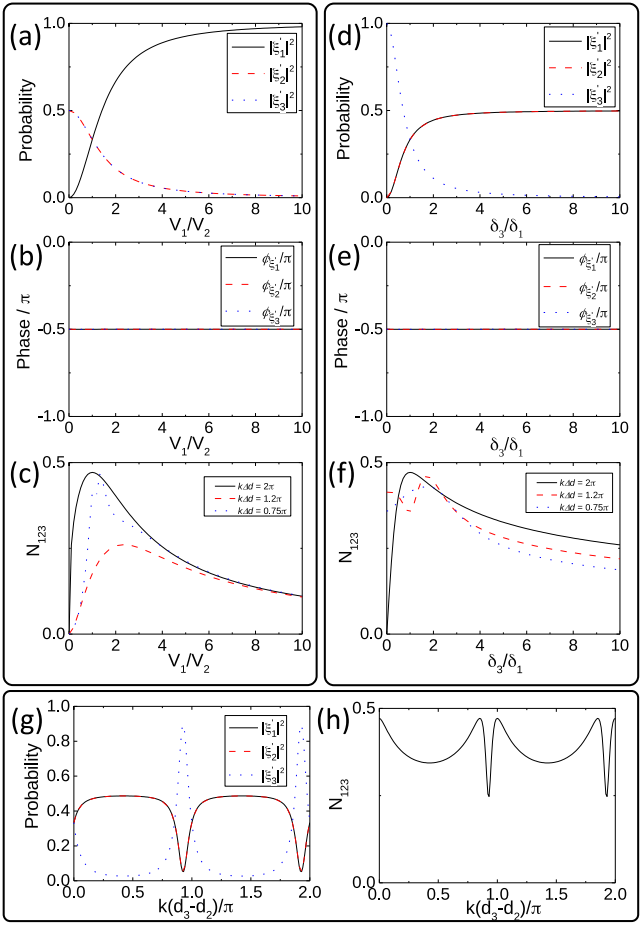}
\caption{\label{fig:3} (a) Excitation probability, (b) phase and (c) tripartite negativity of the W-like state as a function of $V_{1}/V_{2}$ with $V_{2}=V_{3}$, $\delta_{1}=\delta_{2}=\delta_{3}=0.001\Gamma_{\rm wg2}$, $\Gamma_{1}=\Gamma_{2}=\Gamma_{3}=0$ and $k(d_{2}-d_{1})=k(d_{3}-d_{2})=2 m \pi$, where $m$ is an integer. (d) Excitation probability, (e) phase and (f) tripartite negativity of the W-like state as a function of $\delta_{3}/\delta_{1}$ with $V_{1}=V_{2}=V_{3}$, $\delta_{1}=\delta_{2}=0.001\Gamma_{\rm wg1}$, $\Gamma_{1}=\Gamma_{2}=\Gamma_{3}=0$ and $k(d_{2}-d_{1})=k(d_{3}-d_{2})=2 m \pi$. The probability and tripartite negativity when the spacings between quantum dots are different are shown in (g) and (h), respectively.}
\end{figure}

The W-like state can also be obtained by controlling the excitation probability amplitude $\xi'_j$ in Eq.~(\ref{eq:multipartite}). This can be achieved by tuning the coupling strength and detuning as indicated by Eq.~(\ref{eq:relations}). As an example, we show in Fig.~\ref{fig:3}(a) and Fig.~\ref{fig:3}(b) that the ratio of $\xi_1$ and $\xi_2$ (or $\xi_3$) can be controlled by tuning $V_{1}/V_{2}$, where the coupling strengths of QD$_2$ and QD$_3$ are chosen to be equal ($V_2=V_3$). Note that the phases of all QDs remain equal and do not vary with the coupling strength. As another example, Fig.~\ref{fig:3}(d) and Fig.~\ref{fig:3}(e) show that the ratio of $\xi_1$ (or $\xi_2$) and $\xi_3$ can also be controlled by tuning $\delta_{3}/\delta_{1}$, where we take $V_{1}=V_{2}=V_3$ and $\delta_{1}=\delta_{2}=0.001 \Gamma_{\rm wg1}$. Again, the relative phases in Eq.~(\ref{eq:multipartite}) are not affected by manipulating the probability amplitudes. 

One may wonder whether the tripartite entanglement survives or not after these manipulations. To verify that, we calculate the tripartite negativity \cite{Sabin08} in Fig.~\ref{fig:3}(c) (when $V_{1}/V_{2}$ is tuned) and Fig.~\ref{fig:3}(f) (when $\delta_{3}/\delta_{1}$ is tuned). The tripartite negativity $N_{123} = (N_1 N_2 N_3)^{1/3}$ of a tripartite state (note that $N_{123}$ is taken to be $2(N_1 N_2 N_3)^{1/3}$ in Ref.~\cite{Miranowicz04}) is nonzero if the state is entangled. Here, $N_i$ is the negativity \cite{Vidal02}---an entanglement measure---between qubit \textit{i} and the subsystem composed of the remaining qubits. To obtain $N_i$, we take the partial transpose of the density operator $\rho$ of the full system with respect to qubit \textit{i}. If the reduced density operator $\rho^{T_i}$ has negative eigenvalues $\lambda_m$, the negativity is given by $N_i=\sum_{m}|\lambda_m|$; otherwise, $N_i$ is zero. Fig.~\ref{fig:3}(c) and Fig.~\ref{fig:3}(f) show that the tripartite negativity is nonzero over the tuning range of $V_{1}/V_{2}$ or $\delta_{3}/\delta_{1}$ and reaches the maximum when the tripartite state becomes a W state, thereby verifying the preservation of the tripartite entanglement. We note that Fig.~\ref{fig:3}(f) as well as Fig.~\ref{fig:3}(d) and Fig.~\ref{fig:3}(e) can also be viewed as if one QD is intrinsically detuned from the others, which is common due to the inhomogeneity of QDs. The excitation probability of the detuned QD clearly declines as its detuning increases. However, the tripartite negativity remains nonzero even if the excitation probability becomes very low. Experimentally, the inhomogeneity (or the size distribution of the quantum dots) can be reduced by controlling the temperature gradient during the synthesis. For example, a size distribution with a variation of only 2.5--5\% was achieved in \cite{Zlateva07}. To control the emission wavelengths, tuning range of 5 nm has been observed with the Stark effect \cite{Empedocles97,Park07} and tens nm with the strain-induced effect \cite{Veilleux10}.

The ability to control the coefficient weighting in Eq.~(\ref{eq:multipartite}) offers the possibility of generating interesting W-like states for quantum information applications. For example, the W-like state
\begin{equation}
\ket{W_{\rm clone}}=\frac{1}{\sqrt{6}}(2\ket{100}_{\rm ABC}-\ket{010}_{\rm ABC}-\ket{001}_{\rm ABC})
\label{eq:cloning_W}
\end{equation}
as shared by three parties Alice, Bob and Carlie, was proposed by Bur\ss \textit{et al.} to realize the nonlocal cloning of quantum states via teleportation \cite{Burb98}. It works as follows: Alice first performs Bell measurement on her qubit and the state to be cloned. She then broadcasts the measurement outcome to Bob and Charlie. With this knowledge, Bob and Charlie apply appropriate unitary operations on their qubits to make a nearly perfect copy of the state. To prepare such W-like state in our proposed system, we choose the coupling strength $V_{1}/V_{2}=2$ with $V_2=V_3$, $\delta_1=\delta_2=\delta_3=0.001\Gamma_{\rm wg2}$, $\Gamma_1=\Gamma_2=\Gamma_3=0$ and $k(d_2-d_1)=k(d_3-d_2)=2 m \pi$. The following tripartite state can be obtained,
\begin{equation}
\ket{W'}=\frac{1}{\sqrt{6}}(2\ket{e_{1},g_{2},g_{3}}+\ket{g_{1},g_{2},e_{3}}+\ket{g_{1},g_{2},e_{3}}).
\end{equation}
By sharing this W-like state with Alice, Bob and Charlie can clone the target state by performing an additional $\sigma_z$ operation on their own qubits.

\begin{figure}[ht!]
\centering
\includegraphics[width=8.5cm]{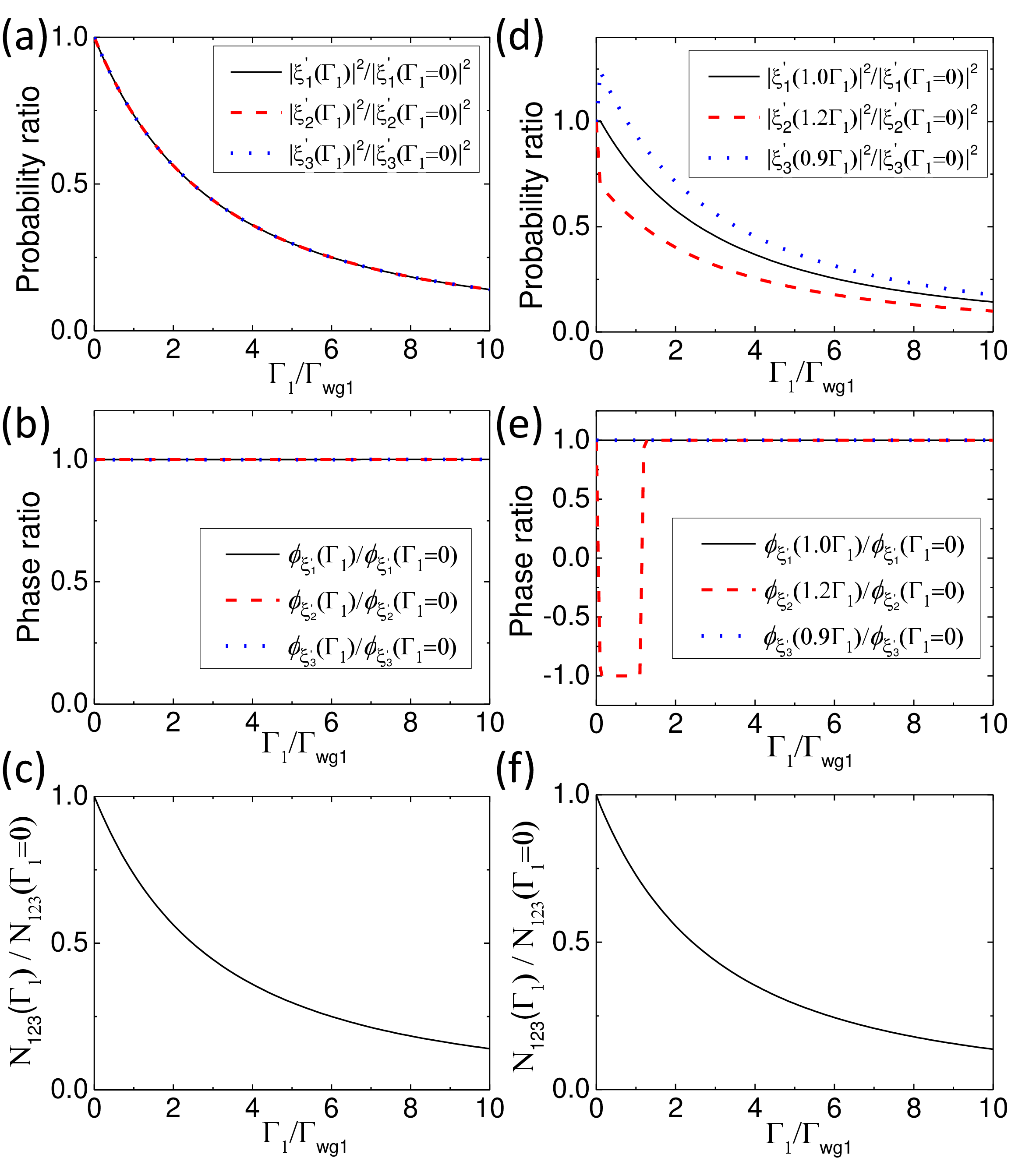}
\caption{\label{fig:4} (a) Probability, (b) phase and (c) tripartite negativity of the W state with equal dissipation as a function of $\Gamma_{1}/\Gamma_{\rm wg1}$ $V_{1}=V_{2}=V_{3}$, $\delta_{1}=\delta_{2}=\delta_{3}=0.001\Gamma_{\rm wg1}$, $\Gamma_{1}=\Gamma_{2}=\Gamma_{3}$, nd $k(d_{2}-d_{1})=k(d_{3}-d_{2})=2 m \pi$. (d) Probability, (e) phase and (f) tripartite negativity of the W-like state with unequal dissipation as a function of $\Gamma_{1}/\Gamma_{\rm wg1}$. In the calculation, $V_{1}=V_{2}=V_{3}$, $\delta_{1}=\delta_{2}=\delta_{3}=0.001\Gamma_{\rm wg1}$, $\Gamma_{2}=1.2\Gamma_{1}$, $\Gamma_{3}=0.9\Gamma_{1}$ and $k(d_{2}-d_{1})=k(d_{3}-d_{2})=2 m \pi$.}
\end{figure}

The dissipation is ignored so far. To see how the dissipation affects the multipartite states, we first consider three QDs with equal decay rates, i.e. $\Gamma_1=\Gamma_2=\Gamma_3$. Fig.~\ref{fig:4}(a), Fig.~\ref{fig:4}(b) and Fig.~\ref{fig:4}(c) show the probability amplitudes, phases and tripartite negativity of the multipartite QD state. One can see that the phases remain unchanged and the tripartite entanglement is preserved even if the excitation probabilities of the QDs decay exponentially. For quantum dots coupled to single photons on a plasmonic waveguide, $\Gamma_i/\Gamma_{{\rm wg}i} = 1/P$ \cite{Akimov07} is small at moderate Purcell factor $P$ (for example, $P \approx 3.7$ in Ref.~\cite{Akimov07}), so the generation of W state is still feasible in the presence of dissipation. Nevertheless, the entanglement will eventually vanish on the time scale of the upper state's lifetime or dephasing time. In the application of the W-like state for optimal quantum cloning \cite{Burb98}, where the Bell measurement or all-optical gates \cite{Predojevic15} are required, the dephasing may be reduced by using the dark exciton state \cite{Labeau03} or the resonant pumping \cite{Ates09,Unsleber16,Nawratha19}. In Fig.~\ref{fig:4}(d), Fig.~\ref{fig:4}(e) and Fig.~\ref{fig:4}(f) we consider three QDs with different decay rates. The tripartite entanglement still survives but the excitation probability of each QD is different and the phase of QD$_2$ flips the sign at high dissipation rate. If equal excitation probability is desired, one may tune the coupling strength and detuning of the QDs to control the probability amplitudes. Interestingly, the multipartite entangled state approaches to the desired state at high dissipation rate. In Fig.~\ref{fig:4}(g) and Fig.~\ref{fig:4}(h), we also study the effect of the unequal spacing on the tripartite entanglement. In general, the entanglement still exists but degrades when the distances between quantum dots are different.

\begin{figure}[ht]
\centering
\includegraphics[width=8.5cm]{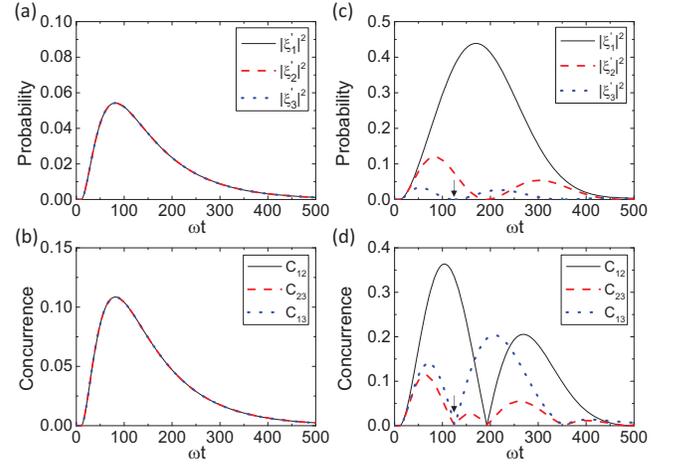}
\caption{\label{fig:5} Excitation probabilities and concurrences when the incident single photon has a waveform of exponential decay. The spacing of the QDs is one wavelength in (a,b) and a quarter of a wavelength in (c,d).}
\end{figure}

\section{Sudden death of entanglement}

One can also control the entanglement of the arbitrary QD pairs in the array. This can be achieved, for example, by temporally shaping the incident single photons. Under such a circumstance, the quantum state at time $t$ will be $\left| \psi (t) \right\rangle = U(t) \left| \psi (0) \right\rangle$, where $U(t) = (2 \pi v_g)^{-1} \int {\rm exp}\left( -iEt/\hbar \right) \left| E \right\rangle \left\langle E \right| dk$, $\left| \psi (0) \right\rangle = \int \psi(x) {\rm exp}(ik_0x)C^{\dagger}_R(x)\left| 0 \right\rangle dx$, and $\psi(x)$ is determined by the single photon's waveform. As an example in Figs.~\ref{fig:5}(a) and \ref{fig:5}(b), we take the waveform to be an exponential decay and the spacing of the QDs to be one wavelength. We then use the Wootters concurrence \cite{Wootters98} to quantify the entanglement of any two QDs. More specifically, by taking the partial trace over the density matrix of the total system, we obtain the reduced density matrix $\rho$ of the two-qubit system and the eigenvalues of the non-Hermitian matrix $\rho\widetilde{\rho}$ in descending order, $\left\{\lambda_1,\lambda_2,\lambda_3,\lambda_4\right\}$, with $\widetilde{\rho}=(\sigma_{y}\otimes\sigma_{y})\rho^{*}(\sigma_{y}\otimes\sigma_{y})$. The concurrence is then calculated by $C(\rho)={\rm max}(0,\sqrt{\lambda_1}-\sqrt{\lambda_2}-\sqrt{\lambda_3}-\sqrt{\lambda_4})$, which reaches the maximal value of 1 when the QD pair is maximally entangled. As shown in Figs.~\ref{fig:5}(a) and \ref{fig:5}(b), we can see that the excitation probability and concurrence of any QD pair both decay exponentially as the waveform of the incident single photon. 

\begin{figure}[ht!]
\centering
\includegraphics[width=8.5cm]{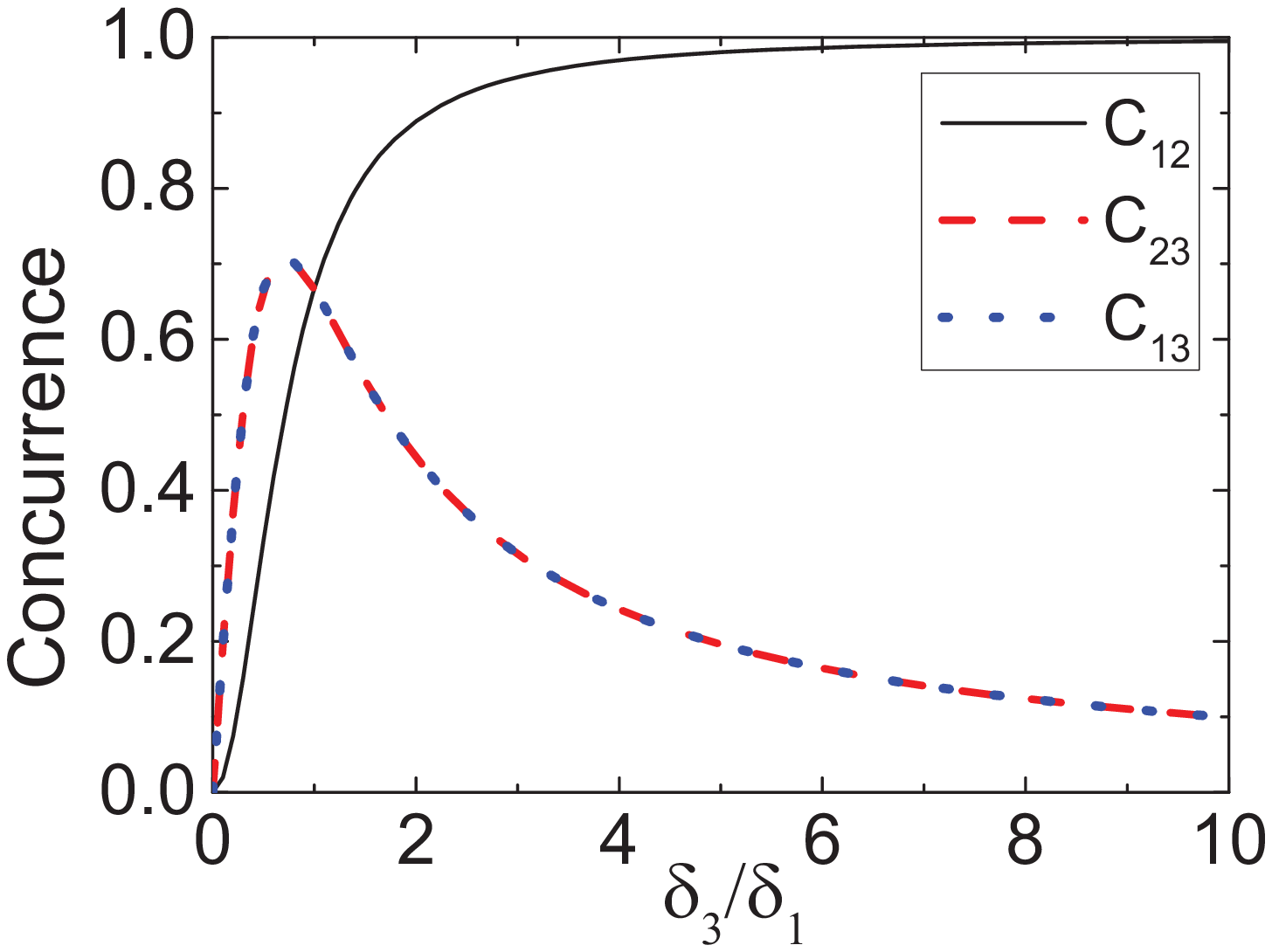}
\caption{\label{fig:6} Concurrences of the entangled QD pairs with $V_{1}=V_{2}=V_{3}$, $\Gamma_{1}=\Gamma_{2}=\Gamma_{3}=0$, and $k(d_{2}-d_{1})=k(d_{3}-d_{2})=2 m \pi$, and $\delta_{1}=\delta_{2}=0.001\Gamma_{\rm wg1}$.}
\end{figure}

An interesting phenomena occurs when the spacing of the QDs is adjusted to a quarter of a wavelength. The two-qubit entanglement then completely disappears at some instants and revives after such sudden death. This can be seen in Figs.~\ref{fig:5}(c) and \ref{fig:5}(d). For example, at the time marked by the arrow, $|\xi_3|^2$, C$_{13}$, and C$_{23}$ all vanish, thus resulting in the sudden death of the entanglement of QD$_1$ and QD$_3$ as well as that of QD$_2$ and QD$_3$. Such death of the entanglement may be protected by engineering the dissipative dynamics \cite{Benito16}. Finally, we note that the entanglement of the arbitrary QD pairs can also be controlled by making the other QDs largely detuned. Fig.~\ref{fig:6} shows the concurrences of different QD pairs as functions of $\delta_{3}/\delta_{1}$ for $\delta_{1}=\pm\delta_{2}$. The maximally entangled states between QD$_1$ and QD$_2$ can be obtained when the detuning of QD$_3$ $|\delta_{3}| > 5|\delta_{1}|$. 

\section{Conclusion}

We have proposed and analyzed the generation of multipartite entanglement in a QD array interacting with a single photon on a plasmonic waveguide. We show that the W state and various W-like states can be generated by controlling the coupling strengths and detunings of the QDs. The generation probability $\textit{P}$ may be optimized by shaping the incident photons. For example, to generate the W state with $k(d_{2}-d_{1})=k(d_{3}-d_{2})=2 m \pi$, $\textit{P} \approx 0.2$ for an exponential-growth-shape incident photon with the 1/\textit{e} time constant equal to $1/3\Gamma_{\rm wg}$ as compared to $\textit{P} \approx 0.1$ for an exponential-decay-shape photon with the same time constant. The multipartite entanglement scheme proposed here can be easily extended to more QDs in one or two dimensional waveguides, with the possibility to increase the entanglement distance by interconnecting each coupled QD-waveguide section by fiber or free-space links. We also show that the entanglement of the arbitrary QD pairs in the array can be generated by dynamically controlled with temporally shaped single photons, in which one may observe the sudden death of entanglement, or manipulating the detunings of the QDs. To prepare the single photons in different waveforms, single photons with long coherence time (for example, from the spontaneous four-wave mixing \cite{Balic05, Du08}, resonant spontaneous parametric down-conversion \cite{Bao08, Scholz09, Wolfgramm11, Chuu12, Wu17}, and cavity quantum electrodynamics \cite{Kuhn02, McKeever04, Thompson06}) may be advantageous for the temporal modulation \cite{Zhang12,Keller04,Kolchin08,Feng17}. We note that the generation of collective states in an atomic ensemble with $\lambda$-type energy levels has been proposed in the DLCZ protocol \cite{Duan01}. This method has also been extended to generate collective atomic states with definite numbers of excitations deterministically or probabilistically using the long-range dissipative or coherent coupling induced via a waveguide \cite{Gonzalez15a,Paulisch17,Gonzalez13,Gonzalez14,Gonzalez15b}. In comparison, ancilla qubits are not required in our work. Finally, we note that the coupling strength used in our calculation, which can be characterized by the Purcell enhancement ($\beta$ factor), ranges from $\beta$ = 9\% to 100\%, which is achievable by the current technology of the photonic-crystal, gap-plasmon, or hybrid plasmonic waveguides (for example, $\beta$ is 98\% in \cite{Arcari14}, 82\% in \cite{Kumar18}, and 73\% in \cite{Kumar19}). 

%the generation of the W states has been proposed or demonstrated with photons, ions, and Rydberg atoms \cite{Yamamoto02,Haffner05,Grafe14,Jo20}. The generation of the collective states with definite numbers of excitations has also been proposed in $\lambda$-type-level atomic ensemble using ancilla qubits \cite{Gonzalez15a,Paulisch17,Gonzalez13,Gonzalez14,Gonzalez15b}. 

\section{Acknowledgments}
The authors thank C.-Y. Cheng for the insightful discussion. This work is supported by the Ministry of Science and Technology, Taiwan (107-2112-M-007-004-MY3, 107-2627-E-008-001, and 107-2745-M-007-001). 

%%%%%%%%%%%%%%%%%%%%%%% References %%%%%%%%%%%%%%%%%%%%%%%%%

%%%%%%%%%% If using BibTeX:

%%%%%%%%%% If preparing manually:
% \begin{thebibliography}{1}
% \newcommand{\enquote}[1]{``#1''}

% \bibitem{Zhang:14}
% Y.~Zhang, S.~Qiao, L.~Sun, Q.~W. Shi, W.~Huang, L.~Li, and Z.~Yang,
%   \enquote{Photoinduced active terahertz metamaterials with nanostructured
%   vanadium dioxide film deposited by sol-gel method,}
%   {\protect\JournalTitle{Optics Express}} \textbf{22}, 11070--11078 (2014).

% \bibitem{OSA}
% {Optical Society}, \enquote{{OSA Publishing},}
%   \url{http://www.osapublishing.org}.

% \bibitem{FORSTER2007}
% P.~Forster, V.~Ramaswamy, P.~Artaxo, T.~Bernsten, R.~Betts, D.~Fahey,
%   J.~Haywood, J.~Lean, D.~Lowe, G.~Myhre, J.~Nganga, R.~Prinn, G.~Raga,
%   M.~Schulz, and R.~V. Dorland, \enquote{Changes in atmospheric consituents and
%   in radiative forcing,} in \enquote{Climate Change 2007: The Physical Science
%   Basis. Contribution of Working Group 1 to the Fourth assesment report of
%   Intergovernmental Panel on Climate Change,}  S.~Solomon, D.~Qin, M.~Manning,
%   Z.~Chen, M.~Marquis, K.~B. Averyt, M.~Tignor, and H.~L. Miler, eds.
%   (Cambridge University Press, 2007).

% \end{thebibliography}

\end{document}